# Magnetic anisotropy in van-der-Waals ferromagnet VI$_3$

A. Koriki[1,2], M. Míšek[3], J. Pospíšil[1], M. Kratochvílová[1], K. Carva[1], J. Prokleška[1], P. Doležal[1], J. Kaštil[3], S. Son[4,5,6], J-G. Park[4,5,6], and V. Sechovský[1]

[1] *Charles University, Faculty of Mathematics and Physics, Department of Condensed Matter Physics, Ke Karlovu 5, 121 16 Prague 2, Czech Republic*

[2] *Hokkaido University, Graduate School of Science, Department of Condensed Matter Physics, Kita10, Nishi 8, Kita-ku, Sapporo, 060-0810, Japan*

[3] *Institute of Physics, Academy of Sciences of Czech Republic, v.v.i, Na Slovance 2, 182 21 Prague 8, Czech Republic*

[4] *Center for Quantum Materials, Seoul National University, Seoul 08826, Korea*

[5] *Center for Correlated Electron Systems, Institute for Basic Science, Seoul 08826, Korea*

[6] *Department of Physics and Astronomy, Seoul National University, Seoul 08826, Korea*

**ABSTRACT**

A comprehensive study of magnetocrystalline anisotropy of a layered van-der-Waals ferromagnet VI$_3$ was performed. We measured angular dependences of the torque and magnetization with respect to the direction of the applied magnetic field within the "*ac*" plane perpendicular to and within the basal *ab* plane, respectively. A two-fold butterfly-like signal was detected by magnetization in the perpendicular "*ac*" plane. This signal symmetry remains conserved throughout all magnetic regimes as well as through the known structural transition down to the lowest temperatures. The maximum of the magnetization signal and the resulting magnetization easy axis is significantly tilted from the principal *c* axis by ~40°. The close relation of the magnetocrystalline anisotropy to the crystal structure was documented. In contrast, a two-fold-like angular signal was detected in the paramagnetic region within the *ab* plane in the monoclinic phase, which transforms into a six-fold-like signal below the Curie temperature $T_C$. With further cooling, another six-fold-like signal with an angular shift of ~30° grows approaching $T_{FM}$. Below $T_{FM}$, in the triclinic phase, the original six-fold-like signal vanishes, being replaced by a secondary six-fold-like signal with an angular shift of ~30°.

## I. INTRODUCTION

Magnetic van der Waals (vdW) materials have recently become hot subjects of interest because of their potential use in atomically thin devices for spintronic and optoelectronic functionalities[1-5]. Exploring new chemistry paths to tune their magnetic and optical properties enable significant progress in fabricating heterostructures and ultra-compact devices[6-9] by mechanical exfoliation[10] or well-controlled element deposition[11]. Despite belonging to a well-studied family of transition metal trihalides, VI$_3$ has received significant attention only recently[10, 12-14]. Nevertheless, the first studies pointed out that the magnetism of VI$_3$ is very complicated. This is probably related to the incomplete high spin state $t_{2g}$ shell in V$^{3+}$ ions here, which allows for a Jahn-Teller distortion. The detailed results of $^{51}$V and $^{127}$I NMR spectroscopy complemented by temperature dependencies of specific-heat and magnetization[15] revealed two distinct ferromagnetic (FM) phases, the ground-state one having the critical temperature $T_{FM}$ = 26 K and another existing between $T_{FM}$ and $T_C$ = 49.5 K (notation is taken from Ref.[14]). VI$_3$ behaves at low temperatures as a hard ferromagnet with a high coercive field $\mu_0 H_c$ ~ 1 T at 2 K for the magnetic field applied parallel to the *c*-axis ($H \| c$)[10, 12], while $H_c$ value is considerably lower in the perpendicular direction ($H \| ab$). In fields above 2 T the magnetization *M* in $H \| c$ is proportional to ~1.2 *M* in $H \| ab$. This anisotropy of magnetization persists at least up to 9 T[10, 12].

A crucial property of the vdW materials with considerable application potential is the strong magnetocrystalline anisotropy[2]. In layered materials, two limits of easy-axis are in-plane (XY model) and out-of-plane (Ising model). The anisotropy features of $VI_3$ were investigated by magnetization measurements. These have shown a different value of saturated magnetization (higher along the *c* axis) and a significantly larger low-temperature magnetization loop hysteresis along the *c* axis[12, 16, 17]. The Raman scattering study has also identified an acoustic magnon mode (the spin-wave gap) and a two-magnon mode in the low-temperature FM state. With the help of linear spin-wave theory calculations, the magnetic anisotropy and intralayer exchange strength are estimated as 1.16 and 2.75 meV. The sizeable magnetic anisotropy implies the stability of the FM order in the 2D limit with a high critical field[16].

First-principles calculations have been used to investigate the magnetism and magnetic anisotropy of $VI_3$ theoretically, however, mostly in the single-layer or bilayer limit[18-21], with only a few works considering the bulk form[10, 22]. Overall, magnetocrystalline anisotropy with an easy axis along *c* has been predicted for both bulk and single layer in these works. A high orbital momentum on V was found in a calculation based on the linear augmented plane wave method, which led to a robust magnetic anisotropy dominated by a single-ion contribution of 15.9 meV per formula unit. Based on these findings, it was concluded that $VI_3$ behaves as an Ising-like ferromagnet, similar to $CrI_3$[23]. In a different calculation based on the projector augmented-wave framework the energy difference between the in-plane and out-of-plane magnetization orientation was found to be strongly dependent on both Hubbard *U* as well as strain. Furthermore, this energy difference changes sign at $U \sim 3.5$ eV, close to the expected value of $U$[24].

Currently, there is a controversy concerning the $VI_3$ crystal structure and its conjunction with magnetism. Kong *et al.* have reported a trigonal (*R-3*) structure at room temperature with a structural transition below $T_s = 78$ K[13]. A contradictory result has been reported by Son et al.[12] showing a structural phase transition at $T_s = 79$ K where the crystal symmetry lowers with cooling from the trigonal *P-31c* structure to a monoclinic *C2/c* in which the system becomes ferromagnetic at temperature $T_C = 50$ K. Conversely, Tian *et al.*[10] claim that the $VI_3$ structural phase transition is similar to the structural transition of $CrI_3$[25], i.e., they observed a lowering of the symmetry with heating through $T_s$ from *R-3* to *C2/m*. A detailed crystal structure study by Doležal *et al.* has confirmed the trigonal (*R-3*) crystal structure at room temperature in agreement with T. Kong[13] and found a subsequent symmetry lowering to a triclinic structure at $T_{FM2} = 32$ K, when the ferromagnetic phase FM I transforms to a different ferromagnetic phase FM II. The connection of the structure with the magnetic phase transition in $VI_3$ suggests a considerable role of magnetoelastic interactions in this compound. It is also supported by magnetostriction-induced changes of the monoclinic-structure parameters at $T_C$. Surprisingly, the temperature of structure transition is decreased by magnetic fields applied along the trigonal *c*-axis and is intact by a magnetic field applied within the basal plane. These results indicate that the magnetic field applied within the three-fold axis of the $VI_6$ octahedrons stabilizes the symmetry of the V honeycombs in the basal plane[26].

The direct observation of the $VI_3$ magnetocrystalline anisotropy was revealed by the angular dependence of magnetization at temperature 2 K in various magnetic fields *H* rotated out-of-plane from the *c* axis to *a* (and *b*) axis and within the *ab* plane. The out-of-plane anisotropy exhibits a "butterfly-like" pattern[17], different from the other 2D FM semiconductor $Cr_2Ge_2Te_6$[27]. On the other hand, the angular signal of magnetization within the basal *ab* plane shows a six-fold-like symmetry. However, anisotropy studies so far performed were mostly focused on the lowest temperature magnetic regime below $T_{FM}$. Further, detailed measurements throughout the entire temperature interval covering all detected magnetic regimes and structural phases are missing and highly desirable to understand complex magnetic interactions and anisotropy in the $VI_3$ vdW ferromagnet. The $VI_3$ magnetic structure, exchange interactions and magnetocrystalline anisotropy are much more complicated than earlier studied Cr tri-halides [28-32].

This paper focuses on studying magnetocrystalline anisotropy features of $VI_3$ in a wide temperature interval by angular-dependent magnetic torque and magnetization measurements for the magnetic field applied out-of-plane marked as "*ac*" (a general perpendicular plane to the *ab* basal plane) and within the *ab* basal plane.

## II. EXPERIMENTAL METHODS

The single crystals were prepared by the chemical vapor transport method, as described elsewhere[12]. The angular dependence of the magnetization was measured by MPMS (Quantum Design) devices using thin single crystals with a sample area of $1.4 \times 1.8$ mm$^2$. The sample was placed on a handmade rotator with a rotation axis orthogonal to the external magnetic field. Magnetic torque was measured using a commercial Quantum Design Torque magnetometer chip attached on a 370° rotator installed inside a 9T PPMS apparatus. To measure the angular dependence of torque within the principal *ab* plane with respect to magnetic field orientation, an L-shaped copper element had to be attached to the torque chip (Fig. 1). The sample was fixed with Apiezon grease. The torque magnetometer chip can typically be calibrated. However, because of installed cooper L element, calibration was not possible to finalize, and so we do not focus on the absolute value of the torque but scale it appropriately. The L-element mass contribution was tested by the rotation of the chip in zero magnetic field; the observed signal was negligible compared to the signal in magnetic fields with the attached sample. The magnetic field is rotating within the *ab* plane of the sample for both measurements. Since the magnetic field was applied within the plane of the thin sample, the shape of samples and the demagnetization factor are expected not to have a significant effect on the results. All samples used for measurements were stored under the inert Ar (6N purity) atmosphere in a glovebox. The contact with air was minimized only on the time of the installation of the rotators into the MPMS and PPMS instruments.

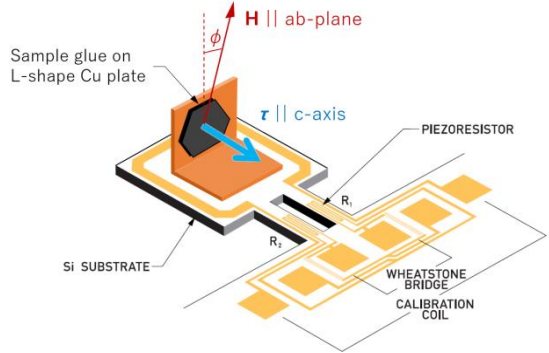

FIG. 1. The scheme of the magnetic torque chip with the installed sample attached to the copper L-element. The scheme was taken from the Quantum Design PPMS manual and modified.

## III. RESULTS

Within our study, we have performed a series of angular-dependent magnetization scans within the *"ac"* plane, perpendicular to the *ab* plane (Fig. 2). Our low-temperature (2 K) and high-magnetic-field (5 T) angular magnetization results in the form of a butterfly-like signal in polar projection agreeing with the previous result in Ref.[17]. Both, the symmetry and positions of the maxima of the angular magnetization signal rotated within the *"ac"* plane remain conserved throughout all magnetic regimes and the structural transition to the triclinic phase[26] at $T_{FM}$. The maximum magnetization signal tilted approximately 40° out of the *c*-axis.

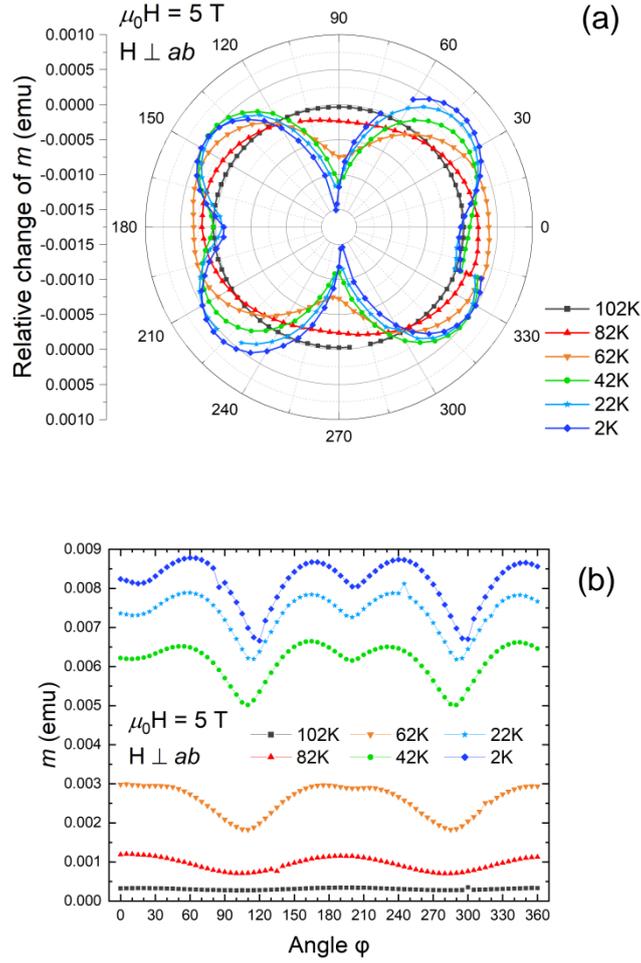

FIG. 2. Angular-dependent magnetization isotherms in magnetic field 5 T rotated within the "*ac*" plane. The polar plot (a) and classical plot (b) shows selected curves at specific magnetic regimes; above $T_C$ (60 K), below $T_C$ (50 K), in the upper vicinity of $T_{FM}$ (30 K), and well below $T_{FM}$ (20 and 10 K).

The angular-dependent magnetic torque measurements within the basal *ab* plane in identical magnetic field 5 T at various temperatures are shown in Fig. 3.

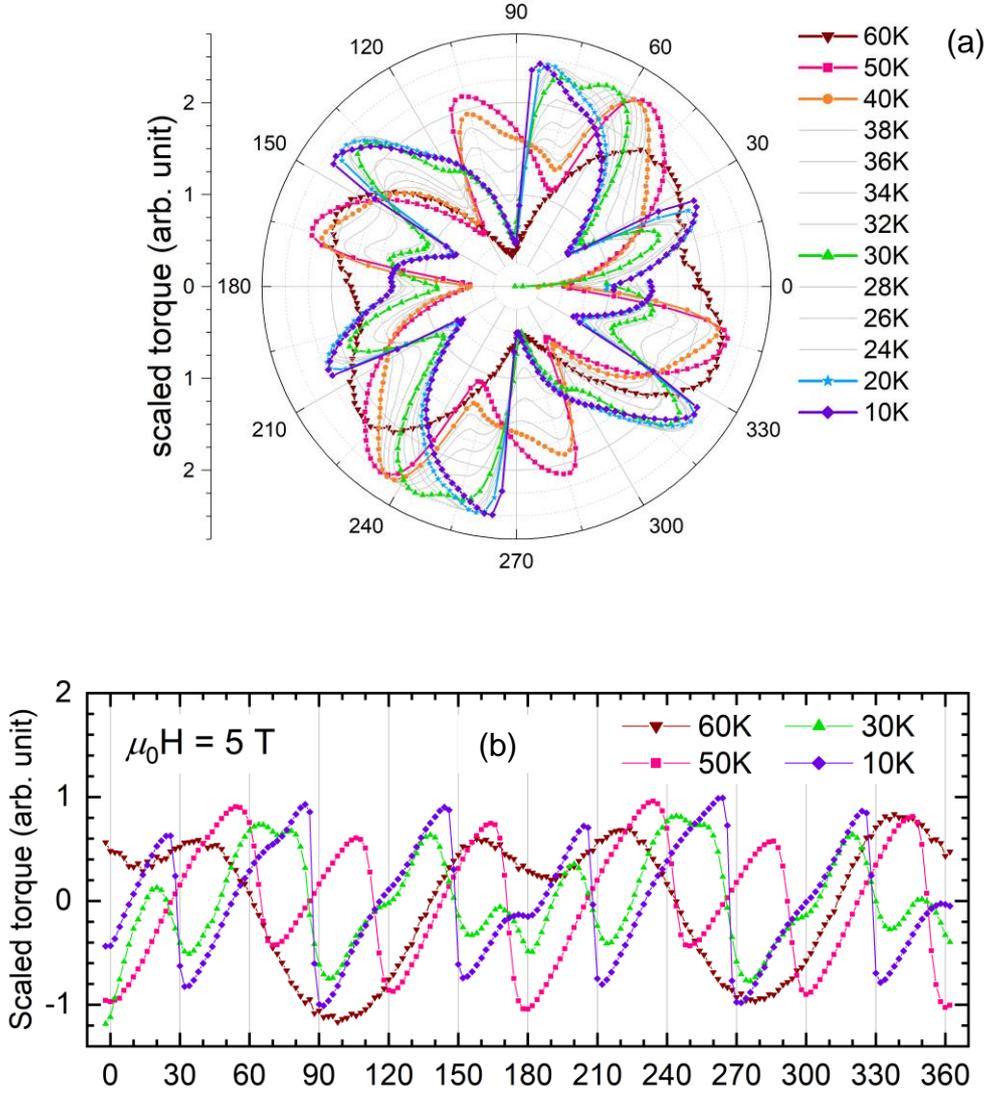

FIG. 3. Angular-dependent magnetic torque scans in magnetic field 5 T rotated within the *ab* plane. The polar plot (a) shows all recorded temperatures. The classical plot (b) shows selected curves at specific magnetic regimes; above $T_C$ (60 K), below $T_C$ (50 K), in the upper vicinity of $T_{FM}$ (30 K), and well below $T_{FM}$ (10 K). The intensity of the torque signal is on a relative scale.

Angular-dependent magnetic torque provides information mainly about the magnetocrystalline anisotropy strength. In contrast, the angular-dependent magnetization measurements show the size of the magnetic moment at the given orientation of the sample with respect to the direction of the external magnetic field. The isothermal scans of the torque have clearly revealed significant evolution of the magnetocrystalline anisotropy (Fig. 3). A double peak-like signal in the paramagnetic region in the monoclinic phase transforms to a six-fold-like signal below $T_C$. The six-fold-like symmetry is present down to the lowest temperatures. However, a detailed analysis of the curves between $T_C$ and $T_{FM}$ shows signatures of the gradual splitting of each maximum when approaching $T_{FM}$. At a temperature of 30 K in the vicinity of $T_{FM}$ the signal is split into two sets of six-fold-like signals of similar intensity mutually rotated by 30° (see Fig. 3b – green line). The original six-fold-like signal, which has emerged at $T_C$, becomes gradually reduced with decreasing temperature in favor of intensity of the second set. The original set vanishes below $T_{FM}$ whereas the second set remains down to the lowest temperature at the fixed angular position.

The angular-dependent scans of magnetization isotherms with magnetic field 5 T applied within the basal *ab* plane are shown in Fig. 4.

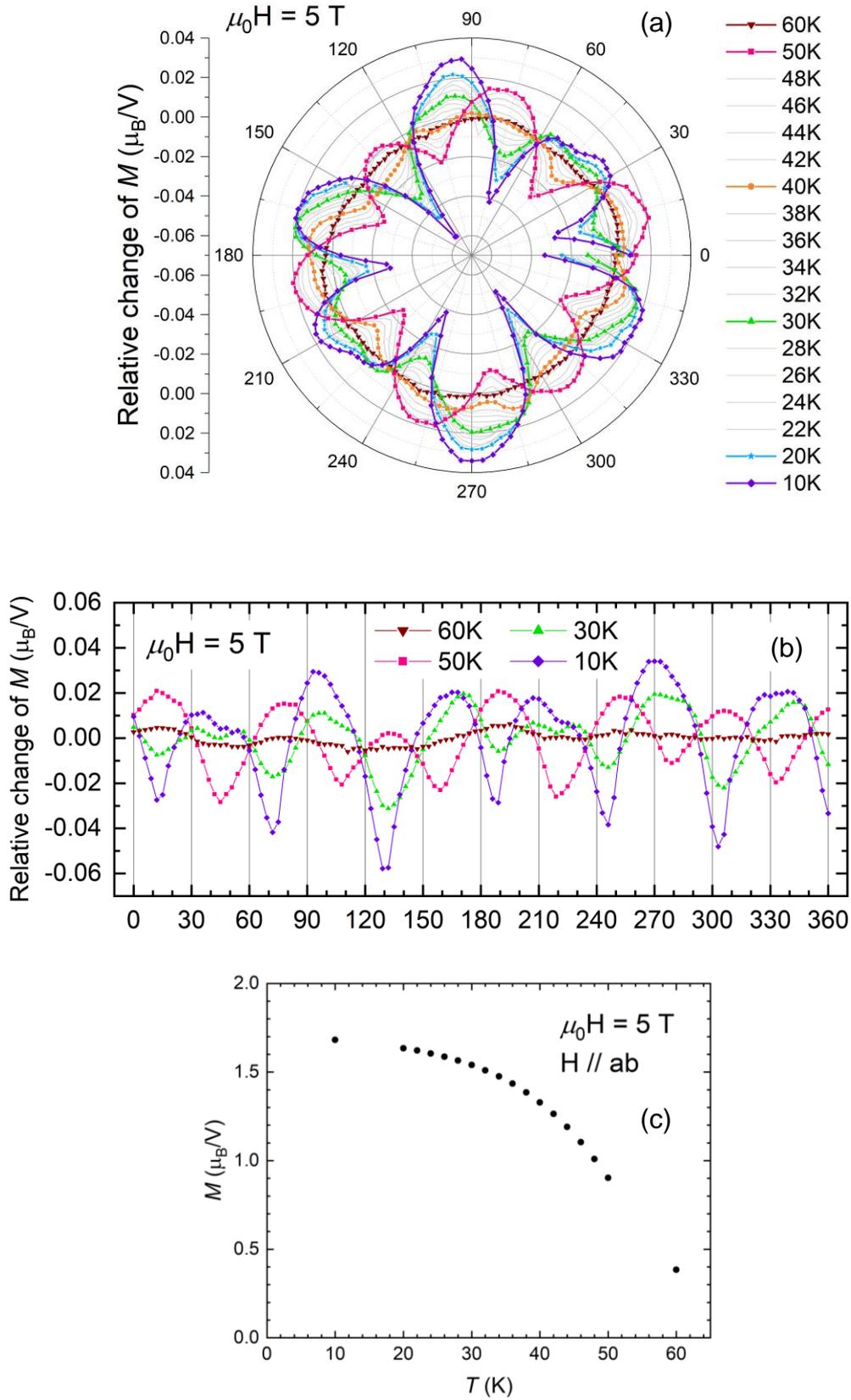

FIG. 4. Angular-dependent magnetization isotherms in magnetic field 5 T rotated within the *ab* plane. The polar plot (a) shows all recorded temperatures. The classical plot (b) shows selected

curves at specific magnetic regimes; above $T_C$ (60 K), below $T_C$ (50 K), in the upper vicinity of $T_{FM}$ (30 K), and well below $T_{FM}$ (20 and 10 K). Panel (c) shows the temperature evolution of the magnetization averaged over the entire angular scan.

The angular-dependent magnetization measurements have shown complementary results with the magnetic-torque measurements. One six-fold-like signal was detected below $T_C$. With further cooling, a clear signature of a second six-fold-like signal rotated by ~30° is observed (see green line in Fig. 4b). The second signal gradually grows at the expense of the original one's intensity. Below $T_{FM}$, the original six-fold-like signal vanishes and only the second one is detected. The signal in the paramagnetic regime at temperatures below $T_s$ is relatively weak, with two-fold-like broad maxima around ~15° and ~195° (Fig. 4b) which can be distinguished analogically to the symmetry of magnetic torque data. The averaged angular value of the magnetic moment at given temperatures is plotted in Fig. 4c and saturates around value 1.7 $\mu_B$/V, which is lower than expected for the $V^{3+}$ ion.

## IV. DISCUSSION

Our detailed experimental study of magnetocrystalline anisotropy by angular-dependent magnetization has unambiguously proved that the magnetic moment is canted from the $c$-axis by ~ 40°. A somewhat surprising result is the robustness of the anisotropy throughout the entire temperature interval covering both ferromagnetic phases and particularly the structural transition from the monoclinic to likely triclinic phase accompanying the magnetic order-to-order transition at $T_{FM}$.

In contrast to the "$ac$" plane, a more complex angular response of the magnetization and torque were detected in the basal $ab$ plane. In the following discussion, we will place the $ab$ plane in the plane of the VI$_3$ layer and take the direction of the two-fold axis as the $b$ axis.

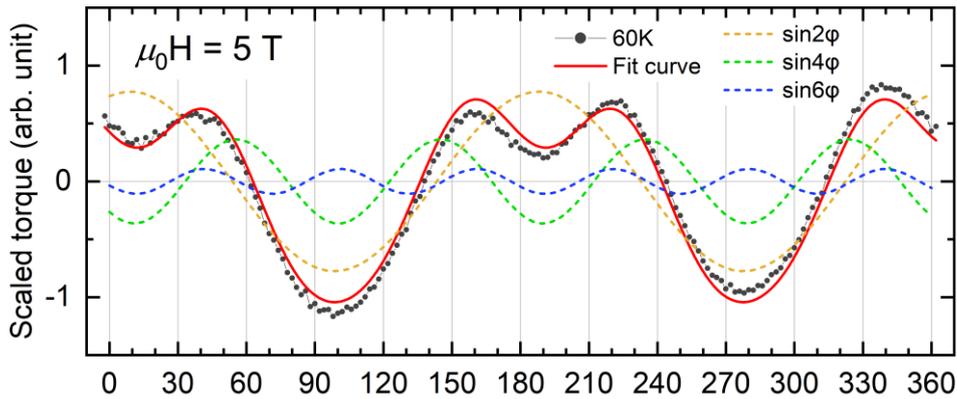

FIG. 5. Angular dependence of torque at 60 K and fitting of the result by combining trigonometric functions.

When the system reveals three-fold rotational symmetries around the $c$-axis, which is the case of the rhombohedral $R$-$3$ space group for VI$_3$ above $T_s$, only the $\sin 6\phi$ term is present in the tensor $\tau_z$ of magnetization torque (Eq. 1, see Appendix).

$$\tau_z = M_x H_y - M_y H_x = \chi_{330} H^6 \sin 6\phi \quad \text{Eq. 1}$$

Focusing on the paramagnetic region at 60 K below $T_s$ (Fig. 5), the best agreement for experimental torque data was found using formula (Eq. 2)

$$\tau(\phi) = k_2 \sin 2\phi + k_4 \sin 4\phi + k_6 \sin 6\phi, \quad \text{Eq. 2}$$

where two-fold and four-fold signals have been observed. This is a direct indication that the symmetry of the crystal below $T_s$ is lowered at least to a monoclinic one at this temperature because the $\sin 2\phi$ and $\sin 4\phi$ contributions are forbidden for three-fold rotational symmetries in the tensor $\tau_z$ (Eq.1) for a rhombohedral structure.

Since the highest point group of a single layer in VI$_3$ is $D_{3d}$, there are three possible orientations of two-fold axes when symmetry is reduced to the monoclinic structure (Fig. 6). These form three domains associated with each other on a three-fold axis perpendicular to the *ab* plane. On the other hand, when a magnetic field is applied to the *ab* plane of a system with two-fold rotations around the *b* (or *a*) axis, the torque in the paramagnetic state can be expressed only by the sum of sin functions (Eq. 2).

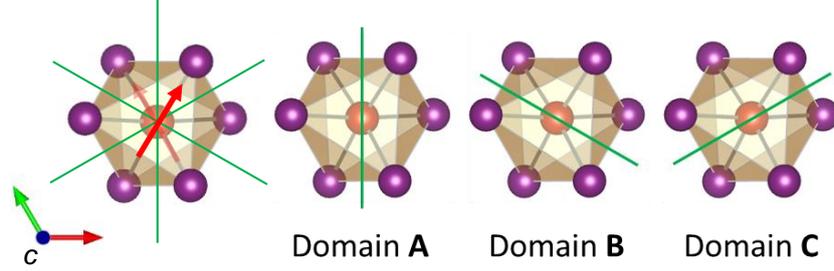

FIG. 6. Schematic view of three two-fold axes and three possible domains when the VI$_3$ transforms from the rhombohedral to the monoclinic structure.

If the volume ratio of the three domains is A : B : C (A + B + C = 1), the torque applied to the entire system is

$$\tau_{\text{tot}}(\phi) = A\tau(\phi) + B\tau(\phi - \pi/3) + C\tau(\phi - 2\pi/3)$$
$$= D\{k_2 \sin 2(\phi + \alpha/2) + k_4 \sin 4(\phi - \alpha/4)\} + k_6 \sin 6\phi \quad \text{Eq. 3}$$

where $D$ and $\alpha$ are constants determined by the volume ratio of the domains. As shown in Fig. 5, the phase of the sine function of the second and fourth oscillation terms is out of phase, which indicates that this is the sum of multiple domains. Since the phase shift due to the domain ratio does not occur in the six-fold oscillation term, the angle at which the torque becomes zero corresponds directly to the main axis of the crystal.

As shown in Fig. 3, at 50 K near the ferromagnetic transition at $T_C$, a six-fold saw-tooth-like signal was observed. This is due to a rapid change in sign due to switching of the ferromagnetic domains, indicating that the magnetization is not fully polarized in the direction of the external magnetic field. This implies that the magnetocrystalline energy due to the VI$_6$ octahedron is sufficiently large with respect to the Zeeman energy. The result can be understood qualitatively by considering the VI$_6$ octahedron to be a regular octahedron as a first approximation. The magnetocrystalline anisotropy energy $E_a^c(\theta, \phi)$, represented by the symmetric representation of the point group $O_h$, can be written as follows

$$E_a^c(\theta, \phi) = K_1 \left\{ \frac{(7\cos^4\theta - 6\cos^2\theta + 3)}{12} - \frac{\sqrt{2}}{3} \sin^3\theta \cos\theta \cos 3\phi \right\} \ldots \text{Eq. 4}$$

This equation is obtained from the commonly used magnetic anisotropy equation $K_1 (\alpha_x^2\alpha_y^2 + \alpha_y^2\alpha_z^2 + \alpha_z^2\alpha_x^2)$ where $\alpha_i$ represents the direction cosine. We obtain this equation by making the [111] axis into a new *c*-axis, which corresponds to the three-fold rotation axis of the VI$_6$ octahedron by an orthogonal transformation. This formula confirms that the easy-axis direction of spontaneous magnetization has six values, $\varphi = 0, \pi/3, 2\pi/3, \pi, 4\pi/3, 5\pi/3$, independent of the sign of $K_1$. Therefore, when the magnetic field is rotated within the *ab* plane, the moment switches to the most stable direction among these easy axes that exist every 60°, as shown in Fig. 7.

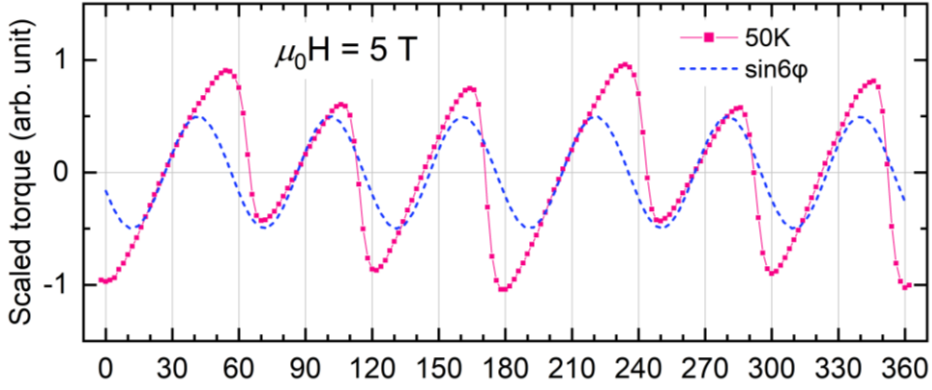

FIG. 7. The angular dependence of torque at 50 K. The dashed blue line is a scaled graph of the six-fold oscillation term from the 60 K result (Fig. 5).

When the magnetic field is in the middle of the two easy axes, an abrupt change in the sign of the torque takes place. Therefore, the main axis of the crystal is also considered to be in this position. In Fig. 7, the six oscillations in the fitting results for the paramagnetic state are shown together, and the two results correspond to each other.

As the temperature was lowered further within the ferromagnetic region, magnetic torque measurements indicate there was a significant change in the easy magnetization axis (Fig. 3). Although this change does not have a clear boundary temperature, another component with an easy axis tilted about ~30° gradually grows and finally replaces the easy axis direction at the lowest temperature. This change cannot be explained, for example, by a change in the sign of the magnetic anisotropy energy $E_a^c(\theta, \phi)$. Previous reports have suggested the presence of a first-order structural phase transition from monoclinic to triclinic at around 30 K[26], and we expect to see a qualitative change reflecting this observation. Detailed crystal structure analysis at low temperatures below $T_{FM}$ is highly desirable. The surprising result is that the "*ac*" plane signal is insensitive to signal symmetry change within the *ab* plane. We can only speculate that the FM I and FM II magnetic structures are very similar, only mutually rotated by 30°.

The results of the angular dependence of magnetization in Fig. 4 are in qualitative agreement with the torque result; we obtained six-fold-like signals. The folding back when the magnetization is reduced corresponds to ferromagnetic domain switching. The original easy axis changes with decreasing temperature about ~30° below $T_{FM}$, which remain conserved down to the lowest temperature of 10 K. For simplicity, to analyze the data, we have decomposed the model into spontaneous magnetization $M_s$, which is independent of the magnitude of the magnetic field, and polarized magnetization $M_p$, which follows the magnetic field. In this model, the result is fitted with a function of |cosφ|+const. in Eq. 5, as shown in Fig. 8.

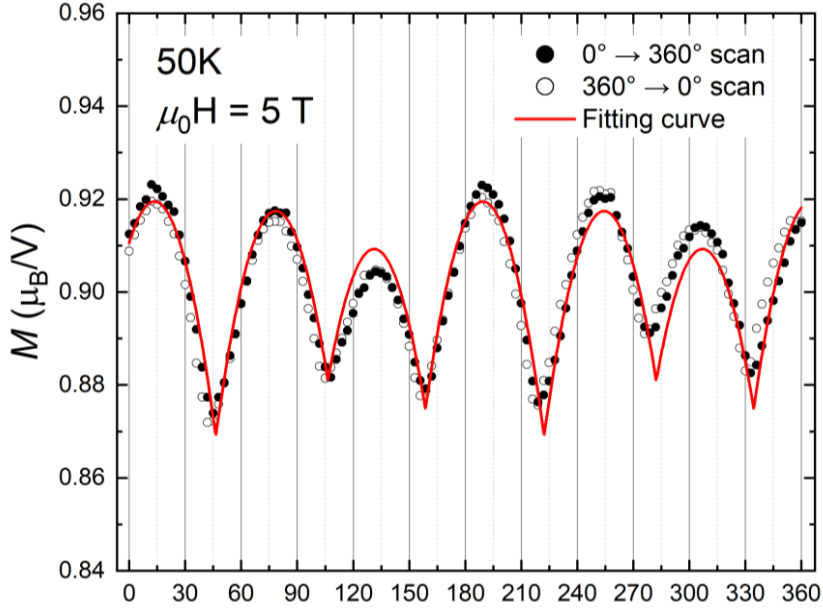

FIG. 8. Angular-dependent magnetization isotherm at 50 K and in magnetic field 5 T rotated within the *ab* plane and the fit at 50 K given by the Eq. 5.

$$M_s \left\{ A \left| \cos\left(\frac{\pi}{w}(\phi - \phi_0)\right) \right| + B \left| \cos\left(\frac{\pi}{w}(\phi - \phi_0 - 60)\right) \right| + C \left| \cos\left(\frac{\pi}{w}(\phi - \phi_0 - 120)\right) \right| \right\} + M_p$$
(Eq. 5)

Fitting result $M_s = 0.44(1)\ \mu_B/V$ and $M_p = 0.62(1)\ \mu_B/V$ is roughly consistent with previous magnetization results of previous experiments at 50 K.

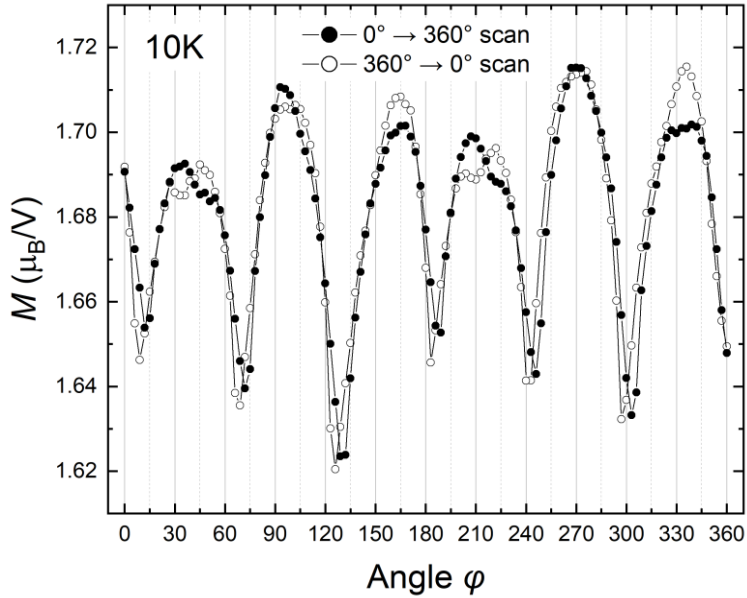

FIG. 9. Angular dependence of the magnetization at 10 K rotated in a clockwise and anticlockwise direction in an identical magnetic field.

Fig. 9 shows the angular dependence of the magnetization at 10 K. The hysteresis is not noticeable at 50 K, but some non-negligible hysteresis is observed at 10 K. Hysteresis, such as the intersection of clockwise and anticlockwise data that exists around 30° and 210°, is typically expected to occur at the minima of magnetization where the domain switches. This suggests that the two components, one developed at around 50 K and the other developed at a lower temperature, coexist as independent signals.

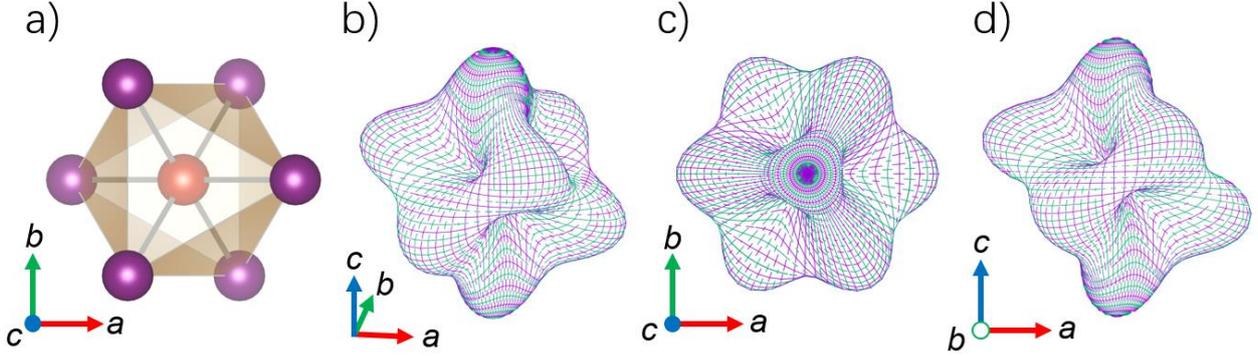

FIG. 10. The projections of the magnetocrystalline anisotropy energy $E_a^c(\theta,\phi)$ for anisotropy in the simplified $O_h$ representation. Pictures a), b), and c) for the principal axis 111 and magnetic moment is stable in the 100 direction; d) projection of easy axes to the *ac* plane.

The observed anisotropy within the *ab* plane can be understood using projections of the magnetocrystalline anisotropy energy $E_a^c(\theta,\phi)$ in a simplified $O_h$ representation for the 111 direction as the principal axis of the $VI_6$ octahedron for magnetic moment stable in the 100 direction of the cube, i.e. the direction of Iodine that leads to the six-fold-like signal (see Fig. 10a, 10c). On the other hand, the projection of the easy axis in the "*ac*" plane (Fig. 10d) leads to a two-fold-like signal, in which the expected moment flipping position is not every 90°.

## V. CONCLUSIONS

We have performed a detailed study of the magnetocrystalline anisotropy of $VI_3$ by angular-dependent torque magnetometry and magnetization within the *ab* plane and the perpendicular "*ac*" plane throughout the entire 10K-60K temperature interval covering both the two ferromagnetic phases and the paramagnetic phase. Irrespective of the temperature change, the symmetry of the two-fold butterfly-like angular signal within the "*ac*" plane remains conserved throughout both FM phases. On the other hand, substantial development of the signal was detected within the *ab* plane. The double peak-like signal in the paramagnetic region in the monoclinic phase transforms to the six-fold-like signal below $T_C$. The original six-fold-like signal appearing at $T_C$ disappears below $T_{FM}$ and the second set of six-fold-like signal rotated by about 30° survives down to the lowest temperatures. The gradual emergence of the secondary six-fold-like signal growth between $T_C$ and $T_{FM}$ and the angular shift of magnetization maxima about ~30° below $T_{FM}$ is the subject of further research. Since the torque contains a periodic function of $2\phi$ and $4\phi$, we corroborate that the three-fold symmetry present in the rhombohedral lattice is broken in the studied temperature range.

Our findings are in agreement with the presence of low symmetry monoclinic and triclinic phases at low temperatures. We have shown that the easy axis is not perpendicular to the layers, but canted by around 40° from the *c*-axis in both FM phases. $VI_3$ is thus not a simple Ising-like ferromagnet. For further progress in this field, it is essential to have detailed knowledge about the predicted triclinic crystal structure below $T_{FM}$ and a neutron diffraction study to reveal the $VI_3$ magnetic structure.

## ACKNOWLEDGMENTS

This work is part of the research program GACR 19-16389J which is financed by the Czech Science Foundation. Work at the Center for Quantum Materials was supported by the Leading Researcher Program of the National Research Foundation of Korea (Grant No. 2020R1A3B2079375) with partial funding by the grant IBS-R009-G1 provided by the Institute for Basic Science of the Republic of Korea. Experiments were performed in the Materials Growth and Measurement Laboratory (see: http://mgml.eu) which is supported within the program of Czech Research Infrastructures (project no. LM2018096). We are indebted to Ross H. Colman for making language corrections.

**APPENDIX: Torque expression in the paramagnetic state**

Considering the situation where time-reversal symmetry is preserved, we expand the magnetization as a function of the magnetic field as follows.

$$M_i = \chi_{ij}^{(2)} H_j + \chi_{ijkl}^{(4)} H_j H_k H_l + \chi_{ijklmn}^{(6)} H_j H_k H_l H_m H_n + \cdots (i, j, k = x, y \text{ or } z)$$

Each $\chi^{(k)}$ in this formula represents the magnetic susceptibility tensors of the *k*-th rank, and here, the general sum reduction rule is used. Since these tensors are symmetrical with respect to the interchange of indices, the distinction between independent components is determined only by the number of the subscripts x, y, and z. Therefore, for simplicity, we have decided to denote these tensors by the number of its xyz, for example, as follows.

$$\chi_{yz}^{(2)} \equiv \chi_{011}, \qquad \chi_{xyyz}^{(4)} \equiv \chi_{121}, \qquad \chi_{xxxxxx}^{(6)} \equiv \chi_{600}$$

We consider the situation where a magnetic field is applied in the *ab*(*xy*) plane, so ($H_x$, $H_y$, $H_z$) = ($H\cos\varphi$, $H\sin\varphi$, 0). Then the general equation for the torque up to the 6th order can be written as follows.

$$\begin{aligned}
\tau_z = &\left\{\frac{1}{2}(\chi_{200} - \chi_{020})H^2 + \frac{1}{4}(\chi_{400} - \chi_{040})H^4 + \frac{5}{32}(\chi_{600} + \chi_{420} - \chi_{240} - \chi_{060})H^6\right\}\sin 2\phi \\
&- \left\{\chi_{110}H^2 + \frac{1}{2}(\chi_{310} + \chi_{130})H^4 + \frac{5}{16}(\chi_{510} + 2\chi_{330} + \chi_{150})H^6\right\}\cos 2\phi \\
&+ \left\{\frac{1}{8}(\chi_{400} - 6\chi_{220} + \chi_{040})H^4 + \frac{1}{8}(\chi_{600} - 5\chi_{420} - 5\chi_{240} + \chi_{060})H^6\right\}\sin 4\phi \\
&- \left\{\frac{1}{2}(\chi_{310} - \chi_{130})H^4 + \frac{1}{2}(\chi_{510} - \chi_{150})H^6\right\}\cos 4\phi \\
&+ \left\{\frac{1}{32}(\chi_{600} - 15\chi_{420} + 15\chi_{240} - \chi_{060})H^6\right\}\sin 6\phi - \left\{\frac{1}{16}(3\chi_{510} - 10\chi_{330} + 3\chi_{150})H^6\right\}\cos 6\phi
\end{aligned}$$

Generally, the tensor is limited in its components by the point group symmetry of the crystal. Eq.1 in the discussion considers a situation in which there are three-fold rotational symmetries around the *c*(*z*) axis. In this case, each component of susceptibility has the following relationship to each other.

$$\begin{aligned}
\chi_{200} &= \chi_{020}, \qquad \chi_{110} = 0 \\
\chi_{400} &= 3\chi_{220} = \chi_{040}, \qquad \chi_{310} = \chi_{130} = 0 \\
\chi_{600} &= 5\chi_{420} = 5\chi_{240} = \chi_{060}, \qquad \chi_{510} = -\chi_{330} = \chi_{150}
\end{aligned}$$

Taking this into account, the equation of torque is expressed as follows.
$$\tau_z = \chi_{330} H^6 \cos 6\phi$$